\documentstyle[11pt]{article}

\title{\bf On the diffraction in the perturbative QCD}
\author{M.Braun \\ Department of High Energy physics,
 University of S.Petersburg,\\
198904 S.Petersburg, Russia}
\def\beq{\begin{equation}}
\def\eeq{\end{equation}}
\def\noi{\noindent}
\begin{document}
\maketitle
\medskip
\noi{\bf Abstract.}

In the framework of the hard pomeron model
 it is shown that at large $Q^2$
the diffractive contribution to the structure functions rises as $Q^{0.3362}$
 and
much slowlier than the total contribution rising as $Q$.
\vspace{1 cm}

{\Large\bf SPbU-IP-1998-2}
\newpage
{\bf 1.} The study of the four-gluon system in the high-colour limit
$N\rightarrow\infty$
reveals that the diffractive amplitude turns out to be totally given by
a contribution  from the triple pomeron interaction [1], in agreement
with the colour dipole picture [2]. In the latter approach the triple
pomeron  amplitude has recently been extensively studied in the high
diffractive mass $M$ region, in relation to the diffractive structure
functions [3]. With the c.m. energy squared $s$, $s_1=M^2$ and $s/s_1$
all being asymptotically large, the amplitude factorizes and, apart
from logarithmic factors, becomes similar to the old triple pomeron
contribution in the Regge-Gribov picture with a triple pomeron coupling
\[
\gamma_{3P}(\kappa)=\frac{32\alpha_s^2N}{\kappa}B
\]
where $\kappa$ is the momentum transferred to the target and $B$ is
a known number [4]
\beq
 B=\frac{g_{3P}}{(2\pi)^4}=\frac{7766.679...}{(2\pi)^4}\simeq 4.983...
\eeq
However it is clear that the total contribution from this asymptotic
region is neglegible at high $s$, the bulk of diffraction coming from
the region of lower $M$, $\alpha_s\ln M\sim 1$,  where the initial
pomeron coupled to the projectile is not in its asymptotic regime.
Because of that  the total diffractive contribution (that is, integrated
over $M$) looks much more complicated and even not factorizable in general.

In this note we want to demonstrate that in the particular case of a highly
virtual photon  projectile, relevant for the structure functions
with $Q^2\rightarrow\infty$, the total diffractive contribution nevertheless
factorizes and has a well-defined behaviour with $Q^2$ as $Q^{0.3662}$.
So it rises with $Q^2$ much slowlier than the non-diffractive contribution
given by a single pomeron exhange, which behaves as $Q$. This fact may serve
as an additional signature of the hard pomeron in the diffractive events.
\vspace{1 cm}

{\bf 2.} The diffractive amplitude (or rather its discontinuity in $s_1$)
corresponding to the triple pomeron interaction is given by
\[
D
=-\frac{\alpha_s^5}{\pi^3}N^{2}(N^{2}-1)\frac{s^{2}}{s_{1}}
\int \prod_{i=1}^{3}d^{2}r_{i}\frac{r_1^2\nabla_1^4}{r_2^2r_3^2}\]
\beq\exp (-i\kappa(r_{2}+r_{3})/2)
\phi_{1}(s_{1},0,r_{1})\phi_{2}(s_{2},\kappa,r_{2})
\phi_{2}(s_{2},-\kappa,-r_{3})\delta^2(r_1+r_2+r_3)
\eeq
Here $\phi_i(s,\kappa,r)$, describe the pomerons coupled to the
projectile ($i=1$) and target ($i=2$), with
the energetic variable $s$, momentum
$\kappa$ and the intergluon transverse distance  $r$; $s_2=s/s_1$.
 The pomerons are
joined by a triple pomeron vertex, whose form was obtained in [4] in the
high-colour limit.
We shall be interested in the total diffractive contribution integrated over
the diffractive mass. In our
normalization
\beq
\frac{d\sigma^D}{d\kappa^2}=\int ds_1\frac{D}{16\pi^2s^2}
\eeq

The pomerons $\phi_{1(2)}$ can be obtained by using the Green function
of the BFKL equation for a given total momentum $G_{l}(s,r,r')$. For the
projectile (see [5]):
\beq
\phi_{1}(s,\kappa,r)=\int d^{2}r'(G_{s}(\kappa,r,0)-
G_{s}(\kappa,r,r')\
\rho_{1}(r')
\eeq
Here $\rho_{1}(r)$ is the colour density of the projectile as a function
of the intergluon distance with the colour factor $(1/2)\delta_{ab}$
and $g^{2}$ separated. As mentioned, for the projectile we choose
 a highly  virtual
 photon with $p_{1}^{2}=-Q^{2}\leq 0$, which splits into
$q\bar q$ pairs of different flavours.
For the transverse photon [5]
\beq
\rho_{1}^{(T)}(r)= \frac{e^{2}}{4\pi^3}
\sum_{f=1}^{N_{f}}Z_{f}^{2}\int_{0}^{1}d\alpha
(m_{f}^{2}{\rm K}_{0}^{2}(\epsilon_{f} r)
+(\alpha^{2}+(1-\alpha)^{2})\epsilon_{f}^{2}{\rm K}_{1}^{2}
(\epsilon_{f} r))
\eeq
where $\epsilon_{f}^{2}=Q^{2}\alpha (1-\alpha)+m_{f}^{2}$ and $m_{f}$
and $Z_{f}$ are the mass and charge of the quark of flavour $f$.
For the longitudinal photon
\beq
\rho_{1}^{(L)}(r)= \frac{e^{2}}{\pi^3}Q^2
\sum_{f=1}^{N_{f}}Z_{f}^{2}\int_{0}^{1}d\alpha
\alpha^2(1-\alpha)^2{\rm K}_{0}^{2}(\epsilon_{f} r)
\eeq
For the hadronic target, we assume an expression similar to (4) with a colour
density $\rho_{2}(r)$ non-perturbative and its explicit
form unknown. 
\footnote{If we put (4) into (2) and use explicit expressions
for the Green
functions $G$, we obtain the diffractive amplitude $D$ in the form given in [3]
with an additional factor 1/16. A part of it, 1/8, is due to our rule to
include a factor 2 into the densities $\rho$. The rest factor 1/2, in our
opinion, comes from using in [3] the double dipole density without a
symmetrizing one half.}

Presenting the $\delta$-function in (2) as an integral over
the auxiliary momentum $q$ we rewrite (2) in the form
\beq
D=\frac{\alpha_s^5}{\pi^3}N^{2}(N^{2}-1)\frac{s^{2}}{s_{1}}
\int \frac{d^{2}q}{(2\pi)^2}\chi_{1}(s_{1},0,q+\kappa/2)
\chi_{2}^2(s_{2},\kappa,q)
\eeq
where
\beq
\chi_{1}(s,0,q)=\int d^{2}rr^2\nabla^{4}\phi_{1}(s,0,r)\exp iqr
\eeq
and
\beq
\chi_{2}(s,\kappa,q)=\int d^{2}r r^{-2}\phi_{2}(s,\kappa,r)\exp iqr
\eeq

The two  pomerons coupled to the target are in their asymptotic regime.
So $\chi_2$ can be
found using an asymptotic expression for the pomeron Green function, in the
same way as in [6], to which paper we refer for the details. One obtains
then
\beq
\chi_{2}(s,\kappa,q)=8s^{\Delta}(\pi/\beta\ln s)^{3/2}f(\kappa)J(\kappa,q)
\eeq
where $\Delta=(\alpha_sN/\pi)4\ln 2$ is the pomeron intercept,
$\beta=(\alpha_sN/\pi)14\zeta(3)$,
\beq
J(\kappa,q)=\int \frac{d^{2}p}{2\pi}\frac{1}
{|\kappa/2+p||\kappa/2-p||q+p|}
\eeq
and $f(\kappa)$ desribes the coupling of the lower pomeron to the target
\beq
\int d^{2}Rd^{2}r\frac{\exp(i\kappa R)r\rho_{2}(r)}{|R+r/2||R-r/2|}=
\pi  f(\kappa)
\eeq
At small $\kappa$ $F_2$ diverges logarithmically:
\beq
f\simeq -4(R/N)\ln\kappa,\ \ \kappa\rightarrow 0
\eeq
where
 \beq R=\frac{N}{2}\int d^2r r\rho_2(r)
 \eeq
 has a meaning of the average target dimension.\vspace{1 cm}

{\bf 3.} For the  pomeron coupled to the projectile
  we have to use the exact Green function. Due to azimuthal symmetry
  we can retain only terms with zero
  orbital momentum in it:
\beq
G_{0}(s,r,r')=(1/8)rr'\int_{-\infty}^{\infty}\frac{d\nu s^{\omega(\nu)}}
{(\nu^{2}+1/4)^{2}}(r/r')^{-2i\nu}
\eeq
where
\beq
\omega(\nu)=2(\alpha_sN/2\pi)(\psi(1)-{\rm Re}\psi(1/2+i\nu))
\eeq
If we take this expression, put it into $\chi_1$, Eq. (8) and integrate
first over $s_1$, as indicated in (3), and afterwards over $r$, we obtain
\beq
\int ds_1 s_1^{-1-2\Delta}\chi(s_1,0,q)=-\frac{4\pi}{q}
\int \frac{d^2r'}{(2\pi)^2}r'\rho_1(r') I(q,r')
\eeq
where $I(q,r')$ is the remaining integral over $\nu$:
\beq
I(q,r')=\int d\nu\frac{(qr'/2)^{2i\nu}}
{2\Delta-\omega(\nu)}\frac{\Gamma(1/2-i\nu)}{\Gamma(1/2+i\nu}
\eeq

The integral (18) can be calculated as a sum of residues of the
integrand at
points $\nu=\pm ix_{k}$, $0<x_{1}<x_{2}<...$, at which
\beq
2\Delta-\omega(\nu)=0
\eeq
Residues in the upper semiplane are to be taken if $qr_{1}/2>1$ and those
in the lower semiplane if $qr_{1}/2<1$. Thus we obtain
\beq
I(q,r_{1})=\frac{2\pi^{2}}{\alpha_sN}\sum_{k}c_k^{(\pm)}
(qr_{1}/2)^{\pm 2x_{k}}
\eeq
where 
\beq
c^{(\pm)}_{k}=
\frac{\Gamma(1/2\mp x_{k})/\Gamma(1/2\pm x_{k})}
{\psi'(1/2-x_{k})-\psi'(1/2+x_{k})}
\eeq
and the signs should be chosen to always have $(qr_{1}/2)^{\pm 2x_{k}}<1$.

The first three roots of Eq. (19) are
\beq 
x_{1}=0.3169,\ \ x_{2}=1.3718,\ \ x_{3}=2.3867
\eeq
with the corresponding coefficients $c_{k}^{(\pm)}$
\[
c_{1}^{(+)}=0.1522,\ \  
c_{2}^{(+)}=-0.1407,\ \  
c_{3}^{(+)}=0.03433,\ \ 
\]
\beq
c_{1}^{(-)}=0.007866,\ \  
c_{2}^{(-)}=-0.001802,\ \  
c_{3}^{(-)}=0.004494,\ \ 
\eeq

Returning to Eq. (3) for $d\sigma/d\kappa^2$ as an integral of the
amplitude
$D$ and
putting expressions for $\chi_2$ and the integrated $\chi_2$,
Eqs. (10) and (17), into the latter we obtain
\[
\frac{d\sigma^D}{d\kappa^2}
=32\pi\alpha_s^4N(N^2-1)\frac{s^{2\Delta}}{(\beta\ln s)^3}
f^2(\kappa)\sum_k\int d^2r r\rho_1(r)\]\beq
[c^{(+)}_k(\frac{r}{2})^{2x_k}B^{(+)}_k(r,\kappa)+
c^{(-)}_k(\frac{r}{2})^{-2x_k}B^{(-)}_k(r,\kappa)]
\eeq
where
\beq
B_k^{(\pm)}(r,\kappa)=\int\frac{d^2q}{(2\pi)^2q}q^{\pm 2x_k}
J^2(\kappa,q-\kappa/2)
\theta (\pm(2/r-q))
\eeq

As we observe, in the general case the factorization property is lost:
the integrals $B^{(\pm)}$ depend nontrivially both on the projectile and
target variables. However one can see that this property is restored in
the  limit of high $Q^2$, relevant for the hadronic structure functions.
In fact, in this limit the characteristic values of $r$ are small:
$r\sim1/Q$. Let us study how the integrals $B^{(\pm)}$ behave at small
$r$. In $B^{(\pm)}$ evidently large values of $q$ are essential.
The integrals $J$ behave as $\ln q/q$ at $q\rightarrow\infty$.
This leads to the following behaviour. 
\[B_k^{(+)}(r,\kappa)\sim r^{1-2x_k}\ {\rm if}\ 2x_k>1\ {\rm and}\
\sim const\ {\rm if}\  2x_k<1\]
\[B_k^{(-)}(r,\kappa)\sim r^{1+2x_k}\]
Combining this with other factors depending on $r$ we see that all terms
multiplying $\rho_1$ in the integrand behave as $r^2$ at small $r$, except
the first term with $k=1$, which, due to $2x_1<1$,  behaves as $r^{1+2x_1}$.
Evidently this term gives the dominant contribution in the limit
$Q^2\rightarrow\infty$, when (24) simplifies to
\beq
\frac{d\sigma^D}{d\kappa^2}
=2^{5-2x_1}\pi e^2 Z^2\alpha_s^4N(N^2-1)c^{(+)}_1
\frac{s^{2\Delta}}{(\beta\ln s)^3}
f^2(\kappa)\frac{b_1B_1}{Q^{1+2x_1}\kappa^{1-2x_1}}
\eeq
where   $Z^2=\sum_f Z^2_f$,
\beq
e^2 Z^2b_1=Q^{1+2x_1}\int d^2rr^{1+2x_1}\rho_1(r)
\eeq
(it does not depend on $Q$) and
\beq
B_1=\kappa^{1-2x_1}\int\frac{d^2q}{(2\pi)^2q}q^{2x_1}J^2(\kappa,q-\kappa/2)
\eeq
(it does not depend on $\kappa$). In (28) $J(\kappa,q)$ is given by (11).
With $x_1=0$ (28) gives the number $B$, Eq. (1).
Numerical calculations give
\beq
b_1^{(T)}=0.3145,\ b^{(L)}=0.04377,\ B_1=17.93
\eeq
So the anomalous dimension in (28) raises the effective triple pomeron
coupling more than three times. Indeces $T,L$ refer to the transverse or
longitudinal photon projectile.

As a rtesult, we observe that in the high-$Q$ limit the expression
for $d\sigma^D/d\kappa^2$ fully
factorizes
in the projectile and target. Its dependence on $Q$ and $\kappa$ turns
out to be intermediate between the eikonal and asymptotic triple pomeron
predictions. It vanishes at large $Q$ as $1/Q^{1+2x_1}$, faster than
the single pomeron
exchange and asymptotic triple pomeron ($\sim1/Q$) but not so fast as the
eikonal prediction $1/Q^2$. It is also singular at $\kappa\rightarrow 0$,
but the singularity is weaker than predicted by the asymptotic triple
pomeron. \vspace{1 cm}

{\bf 4.} The diffractive contributions to the structure functions are
obtained from (26) in the standard manner:
\beq
F_2^D(x, Q^2, \kappa)=\frac{Q^2}{\pi e^2}\frac{d\sigma^D}{d\kappa^2},
\ s\rightarrow 1/x
\eeq
(one has to sum the contributions from the transverse and longitudinal
photon projectiles). From (30) one concludes that at large $Q^2$ the
diffractive contribution behaves as $Q^{1-2x_1}=Q^{0.3362}$. This is to be
compared with the non-diffractive contribution given by a single pomeron
exchange in the same approach:
\beq
F_2(x,Q^2)=\frac{11}{32}Z^2\alpha_s^2\frac{N^2-1}{N}
\sqrt{\frac{\pi}{\beta\ln s}}s^{\Delta}QR
\exp\left(-\frac{\ln^2QR}{\beta\ln s}\right)
\eeq
At $\ln^2 QR<<\beta\ln s$ it roughly behaves as $Q$ and so dominates over the
diffractive
contribution at a given $x$ and rising $Q$. However it rises with $1/x$
only as $x^{-\Delta}$ whereas the diffractive contribution rises as its
square (modulo logarithms). So in different regions of the $x,Q^2$ space
the relative magnitude of the diffractive and non-diffractive contributions
may be different.

To have an idea about the numerical magnitude of the diffractive
contribution we have to make some specific assumptions about the hadronic
colour density $\rho_2(r)$ which enters the form -factor $f(\kappa)$,
Eq.(12). In accordance with (14) it shoud be normalized at $\kappa=0$ as
\beq
(N/2)\int d^2r\rho_2(r)=1,\ \ \kappa=0
\eeq
However its form as a function of $r$ and $\kappa$ is unknown. To avoid
introduction of several hadronic scales and also guided by the form of the
photonic density $\rho_1(r)$ at large $Q^2$ and arbitrary $\kappa$ [1],
we choose
\beq
\rho_2(r)=\frac{\alpha^2}{\pi N}e^{-r(\alpha+\kappa)}
\eeq
The single dimensional parameter $\alpha$ is related to the average
target dimension as $\alpha=2/R$. With this choice we obtain
\beq
f(\kappa)=(R/N)\tilde{f}(\kappa/\alpha)
\eeq
where
\beq
\tilde{f}(\kappa)=\int\frac{d^2p}{\pi p|p+\kappa|}
\frac{2(1+|\kappa|)^2-(p+\kappa/2)^2}{((1+|\kappa|)^2+(p+\kappa/2)^2)^{5/2}}
\eeq
With our choice the form factor $f$ goes down with $\kappa$ as a power.
 At $\kappa\rightarrow 0$ it blows up
logarithmically  (see Eq. (13)).

Integrating (30) over all $\kappa$ we get the total diffractive contribution
to the structure function
corresponding to transitions into the proton intermediate states
\beq
F_2^D(x,Q^2)=64 Z^2 \alpha_s^4\frac{N^2-1}{N}b_1B_1c(QR)^{1-2x_1}
 \frac{s^{2\Delta}}{(\beta\ln s)^3}
\eeq
where the number $a$ is given by
\beq
a=\int_0^{\infty}d\kappa \kappa^{2x_1}\tilde{f}^2(\kappa) \simeq 6.033
\eeq
The ratio of the diffractive to non-diffractive contributions results
\beq
F_2^D/F_2=c\alpha_s^2(QR)^{-2x_1} \frac{s^{\Delta}}{(\beta\ln s)^{5/2}}
\exp\left(\frac{\ln^2QR}{\beta\ln s}\right), \ s\rightarrow 1/x
\eeq
with a coefficient $c$ given by
\beq
c=\frac{2048}{11\sqrt{\pi}}ac_1^{(+)}B_1(b_1^{(T)}+b_1^{(L)})
\eeq

In Table we present values of this ratio at $x=10^{-4},\ 10^{-5},\ 10^{-6}$
and  $Q^2=10,\ 100,\ 1000\ (GeV/c)^2$.
We have chosen $\Delta=0.282$ and $1/R=0.522\ GeV $ (see [3], our $R$ is their
$r_0/2$). As one observes the $x$-dependence of the ratio is quite weak in
this region. On the contrary the $Q^2$ dependence is significant:
the relative contribution of the diffractive events goes down by a
factor of two as $Q^2$ rises from 10 to 1000 $(GeV/c)^2$. The total relative
number of diffractive events reaches $\sim$ 10\% at the smallest values
of $x$ and $Q^2$ in the studied interval.\vspace{1 cm}

{\bf References}

\noi 1. M.A.Braun and G.P.Vacca, Bologna univ. preprint, hep-ph/9711486.\\
2. A.Mueller, Nucl. Phys.,{\bf B415} (1994) 373.\\
 R.Peschanski, Phys. Lett. {\bf B409} (1997) 49.\\
3. A.Bialas, H.Navelet and R.Peschanski, Saclay preprints
hep-ph/9711236, hep-ph/9711442.\\
4. G.P.Korchemsky, preprint LPTHE-Orsay-97-62; hep-ph/9711277.\\
5. N.N.Nikolaev and B.G.Zakharov, Z. Phys. {\bf C49} (1991) 607.\\
6. M.A.Braun, Z.Phys. {\bf C71} (1996) 123.\\
\newpage
\begin{center}
{\Large\bf Table}\vspace{1cm}

{\bf Ratio of the diffractive to nondiffractive contributons to $F_2(x,Q^2)$}
\vspace{0.5 cm}

\begin{tabular}{|r|c|c|c|}\hline
$Q^2\ (GeV/c)^2$& $x=10^{-4}$&$x=10^{-5}$&$x=10^{-6}$\\\hline
10   & 0.0750 & 0.0789 & 0.0932 \\\hline
100  & 0.0512 & 0.0502 & 0.0566 \\\hline
1000 & 0.0413 & 0.0366 & 0.0385 \\\hline
\end{tabular}
\end{center}
\end{document}